\documentstyle{article}
\newcommand{\beq}{\begin{equation}}
\newcommand{\eeq}{\end{equation}}
\newcommand{\beqa}{\begin{eqnarray}}
\newcommand{\eeqa}{\end{eqnarray}}
\newcommand{\bea}{\begin{eqnarray}}
\newcommand{\eea}{\end{eqnarray}}

\newcommand   {\r}       {{\bf r}}

\newcommand  {\JCP}      {{\it J.\ Chem.\ Phys.\ }}

\newcommand  {\JP}       {{\it J.\ Phys.\ }}

\newcommand  {\M}        {{\it Macromolecules\ }}

\newcommand  {\MP}       {{\it Molec.\ Phys.\ }}

\newcommand  {\ProSci}   {{\it Protein\ Sci.\ }}

\newcommand  {\PNAS}     {{\it Proc.\ Natl.\ Acad.\ Sci.\ USA\ }}

\newcommand  {\PRL}      {{\it Phys.\ Rev.\ Lett.\ }}
\newcommand  {\Sci}      {{\it Science\ }}

\input{psfig}
\input epsf

\begin{document}

\begin{titlepage}

\begin{flushright}
LU TP 97-33\\
July 9, 1998\\
Revised version\\
\end{flushright}
\vspace{.6in}
\LARGE
\begin{center}
{\bf Monte Carlo Procedure for Protein Design}\\
\vspace{.3in}
\large
Anders Irb\"ack\footnote{irback@thep.lu.se}, 
Carsten Peterson\footnote{carsten@thep.lu.se},\\ 
Frank Potthast\footnote{frank@thep.lu.se} and 
Erik Sandelin\footnote{erik@thep.lu.se} \\
\vspace{0.10in}
Complex Systems Group, Department of Theoretical Physics\\ 
University of Lund,  S\"{o}lvegatan 14A,  S-223 62 Lund, Sweden \\
{\tt http://thep.lu.se/tf2/complex/}
\vspace{0.3in}  

Submitted to {\it Physical Review} {\bf E}

\end{center}

\normalsize

\vspace{0.3in}

Abstract: 

A new method for sequence optimization in protein models is 
presented. The approach, which has inherited its basic philosophy 
from recent work by Deutsch and Kurosky [\PRL {\bf 76}, 323 (1996)]
by maximizing conditional probabilities rather than minimizing energy 
functions, is based upon a novel and very efficient multisequence 
Monte Carlo scheme. By construction, the method ensures that the designed 
sequences represent good folders thermodynamically. A bootstrap procedure for 
the sequence space search is devised making very large chains feasible.
The algorithm is successfully explored on the two-dimensional HP model with 
chain lengths $N$ = 16, 18 and 32. 

\end{titlepage}

The ``inverse'' of protein folding, sequence optimization, is of utmost 
relevance in the context of drug design. This problem, which amounts 
to finding optimal amino acid sequences given a target structure, has also 
been investigated in the context of understanding folding 
properties of coarse-grained models for protein folding. 
Such models are described by energy functions $E(r,\sigma)$, where 
$r=\{\r_1,\r_2,..,\r_N\}$ denotes the amino acid coordinates 
and $\sigma=\{\sigma_1,\sigma_2,..,\sigma_N\}$ the amino acid 
sequence. 

Good folding sequences fold fast and in a stable way into the desired target 
structure. A brute force search for sequences meeting these criteria is 
prohibitively time-consuming even in minimalist models for protein folding. 
Although it has been possible to apply this type of criteria to a simple 
helix-coil model~\cite{Ebeling:95}, it is essential to find more efficient 
strategies. A fairly drastic simplification was proposed in 
\cite{Shakhnovich:93b}, 
where the problem is approached by minimizing $E$ with respect to 
$\sigma$ with $r$ clamped to the target structure, $r_0$. 
This method is very fast since no exploration of the conformational 
space is involved, but, unfortunately, it fails for a number of 
examples (see e.g. \cite{Ebeling:95,Deutsch:95/96,Seno:96}). 
Recently, a more generic scheme was suggested~\cite{Deutsch:95/96}, 
which aims at optimizing the conditional probability $P(r_0|\sigma)$, 
i.e. the Boltzmann weight, rather than $E(r_0,\sigma)$. 
This approach has the advantage that entropy effects 
are taken into account, but its usefulness is not obvious since 
maximizing $P(r_0|\sigma)$ is a non-trivial task. 
In fact, the calculations in \cite{Deutsch:95/96} involved simplifying 
assumptions about both the form of $P(r_0|\sigma)$ and the conformational 
space. In this letter we present a practical 
Monte Carlo (MC) procedure for performing the maximization of 
$P(r_0|\sigma)$.

Thermodynamical characteristics for good folders 
are that the ground state minima are well separated from other 
states --- at finite $T$ the system spends a long time in the 
ground state well. In lattice models, where for relatively small 
chains the states are enumerable, this is often taken as 
non-degeneracy of ground state and that the latter is well separated 
from higher energy states. One expects that working with finite $T$ 
distributions in the matching process singles out those 
optimal sequences that have good folding properties in terms of 
non-degeneracy. 
Indeed, in~\cite{Deutsch:95/96} when exploring the technique on 
lattice models, superior results are obtained when comparing with 
what was obtained in~\cite{Shakhnovich:93b}, where $E(r_0,\sigma)$ was 
minimized.

Computationally, straightforward MC approaches for maximizing  
$P(r_0|\sigma)$ are extremely tedious. Our novel  MC methodology  
is based on the {\it multisequence method} \cite{Irback:95b}, where 
both sequence and coordinate degrees of freedom are subject to 
simultaneous moves. The basic idea is to perform a single simulation 
of a joint probability distribution $P(r,\sigma)$ rather than repeated 
simulations of  $P(r|\sigma)$ for 
different fixed $\sigma$. Hence, our approach is fundamentally 
different from that of~\cite{Seno:96}.

Our method for maximizing $P(r|\sigma)$ is explored on the
two-dimensional HP lattice model~\cite{Lau:89} with chain lengths 
$N=16$, $18$ and $32$. For $N$=$16$ we study an example used in 
\cite{Seno:96,Shakhnovich:93b,Deutsch:95/96}. 
The results for both $N$=$16$ and $18$
are checked against exact enumerations, whereas for $N=32$ we use a 
target structure constructed ``by hand''. Our method 
reproduces the exact results extremely rapidly whenever comparisons 
are feasible. Furthermore, the method has quite some potential to 
deal with very large chains.

\begin{figure}[htbp]
\vspace{1.5in}
\begin{center}
\vspace{-35mm}
\mbox{\psfig{figure=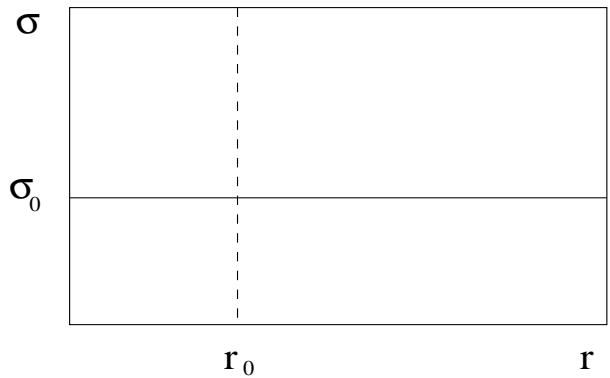,width=8cm,height=5cm}}
\vspace{0mm}
\end{center}
\caption{The ($r,\sigma$)-plane (see text).}
\label{plane}
\end{figure}

The problem of finding thermodynamically optimal sequences 
given a target structure $r_0$ is simple to formulate mathematically 
---  maximize with respect to $\sigma$ the conditional probability 
\beqa
\label{P}
P(r_0|\sigma) = \frac{1}{Z(\sigma)}\exp (-E(r_0,\sigma)/T)\\
\label{Z}
Z(\sigma) ={\sum_r \exp(-E(r,\sigma)/T)}
\eeqa
where Eq.~(\ref{P}) can be rewritten in terms of the
free energy $F(\sigma)=-T\ln Z(\sigma)$ as
\beq
\label{P1}
P(r_0|\sigma) = \exp [-(E(r_0,\sigma)-F(\sigma))/T]
\eeq
Hence, for each $\sigma$ one needs to estimate $P(r_0|\sigma)$ 
which in turn involves a sum over all possible $r$. The situation is 
shown in Fig.~\ref{plane}, where the horizontal line represents 
the region probed in protein folding. In the simplified approach to 
the inverse problem \cite{Shakhnovich:93b}, 
minimizing $E(r_0,\sigma)$, one works along the vertical line.  
Maximizing $P(r_0|\sigma)$ is a real challenge since it requires 
sampling of the entire ($r,\sigma$)-plane.
Refs.~\cite{Deutsch:95/96,Seno:96} approached this problem by using 
simulated annealing in sequence space. The key difficulty then 
is to estimate the partition function $Z(\sigma)$ (Eq.~(\ref{Z})). 
In~\cite{Deutsch:95/96} this was done using the lowest-order
term in the cumulant expansion of $F(\sigma)$. This approximation is   
valid at high temperature but it is unclear how good it is in the 
temperature regime of interest here. Ref.~\cite{Seno:96}, on 
the other hand, used a chain growth MC method to estimate $Z(\sigma)$. 
In this case one has a nested MC, where the inner part by itself is far 
from trivial.   

In~\cite{Deutsch:95/96,Seno:96} these methods were successfully
tested on examples where a simple minimization of $E(r_0,\sigma)$ along 
the vertical line in Fig.~\ref{plane} fails. The chains were short    
enough for the results to be tested against exact enumerations.
Another difference between optimizing  $P(r_0|\sigma)$ versus $E(r_0,\sigma)$ 
is that the latter requires an optimization constrained to a preset net 
hydrophobicity.

In this letter we take a quite different path capitalizing on  
the multisequence method \cite{Irback:95b}. Here the basic strategy is to 
create an enlarged configuration space; the sequence $\sigma$ becomes a 
dynamical variable. Hence, $r$ and $\sigma$ are put on a more equal 
footing, which, in particular, enables us to avoid a nested MC. 

Our starting point is the joint probability distribution 
\beqa
\label{joint}
P(r,\sigma) & = &\frac{1}{Z} \exp(-g(\sigma)-E(r,\sigma)/T)\\
Z & = & \sum_\sigma \exp(-g(\sigma)) Z(\sigma)  
\eeqa
where $\{g(\sigma)\}$ is a set of tunable parameters, which 
govern the marginal distribution  
\beq
\label{marg}
P(\sigma) = \sum_r P(r,\sigma) = 
\frac{1}{Z} \exp(-g(\sigma)) Z(\sigma)
\eeq
From the Bayes relation $P(r,\sigma)=P(r|\sigma)P(\sigma)$ one 
obtains the desired conditional probabilities, Eq.~(\ref{P}), which
of course are independent of the choice of $g(\sigma)$. 

The choice of $g(\sigma)$ is crucial for the efficiency of the method.
At first sight, it may seem that one would need to estimate $Z(\sigma)$
in order to obtain reasonable $g(\sigma)$. However, a convenient choice is
\beq
\label{gs}
g(\sigma) = -E(r_0,\sigma)/T 
\eeq
Maximizing $P(r_0|\sigma)$ in Eq. (\ref{P}) corresponds to minimizing 
the quantity  
\beq
\label{DF}
\Delta F_0 = -T\ln P(r_0|\sigma)= E(r_0,\sigma)-F(\sigma) 
\eeq
which for the choice in Eq. (\ref{gs}) can be rewritten as 
\beq
\label{DF2}
\Delta F_0 = T\ln P(\sigma)+T\ln Z  
\eeq
Hence, neglecting an unimportant constant, $\Delta F_0$ 
can be obtained directly from the marginal distribution $P(\sigma)$.

The joint distribution $P(r,\sigma)$ can be simulated by using 
separate ordinary $r$ and $\sigma$ updates.  
This single simulation of $P(r,\sigma)$ replaces simulations of
$P(r|\sigma)$ for a number of different fixed $\sigma$. This is quite 
convenient. However, it should be stressed that the major motivation for 
using this scheme is its efficiency. In fact, it was 
demonstrated in \cite{Irback:95b} that the simulation of $P(r,\sigma)$ can 
be much faster than the simulation of $P(r|\sigma)$ even for {\it a single} 
$\sigma$; the exploration of the conformational space becomes more efficient
when the sequence is allowed to fluctuate.   

The number of sequences that can be studied in a multisequence simulation
is of course limited. It is therefore desirable to incorporate a step in 
which ``bad'' sequences are removed. This elimination step can be 
formulated in different ways. In our calculations a sequence $\sigma$ is 
removed as soon as some structure $r\ne r_0$ is encountered for 
which $E(r,\sigma)\le E(r_0,\sigma)$. This process is continued until
the remaining sequences can be studied through a final multisequence run.   
If the elimination proceeds for a sufficiently long time, then  
the surviving sequences are, by construction, those that have 
the desired structure as their non-degenerate ground state. It should
be pointed out that the elimination process serves two purposes. In
addition to bringing down the number of sequences to a manageable
level, it also tends to make the distribution $P(\sigma)$ more uniform,
which is instrumental for the efficiency of the final multisequence run.
Note that for a set of sequences all having $r_0$ as their unique
ground state, one has $\exp(-g(\sigma))Z(\sigma)\approx 1$ to leading 
order at low $T$, independent of $\sigma$, which implies that      
$P(\sigma)$ becomes uniform in the zero temperature limit.  


We have performed extensive numerical explorations on the HP 
model~\cite{Lau:89} for a variety of sizes and target structures and 
find that the approach consistently identifies the appropriate sequences 
in an efficient way. In this letter we report on results for $N=16$ and $18$, 
which have been previously used for evaluating design algorithms 
\cite{Deutsch:95/96,Seno:96}. Also results for $N=32$ are presented in 
some detail.

The HP model~\cite{Lau:89} is defined by 
%
\beq
\label{HP}
E(r,\sigma) = 
-\sum_{i<j}\sigma_i\sigma_j\Delta(r_i - r_j)
\eeq
where $\Delta(r_i - r_j)$ = $1$ if $r_i$ and $r_j$ are nearest
neighbor monomers but non-adjacent along the chain and zero otherwise. 
Dependent upon whether $\sigma_i$ is hydrophobic (H) or polar (P), one has 
$\sigma_i$=$1$ and $0$, respectively.
We work on the square lattice, for which it is known that sequences
with non-degenerate ground states are not too rare~\cite{Dill:95}.
All simulations are performed using standard Metropolis 
steps~\cite{Metropolis:53} in $\sigma$. In $r$ we use three types
of elementary moves: one-bead, two-bead and pivot~\cite{Lal:69,Sokal:95}.
A sweep refers to a combination of these three move types followed
by one attempt to update $\sigma$.

We first test our method for the $N=16$ target structure studied 
in~\cite{Seno:96} (see Fig.~1 in~\cite{Seno:96}). There is
one sequence which has this structure as its non-degenerate ground
state, as can be shown by exact enumeration. It turns out that the
design procedure in~\cite{Seno:96} is able to find this 
sequence, while the methods in~\cite{Shakhnovich:93b,Deutsch:95/96} 
fail to do so. Our calculation is carried out starting from the set 
of all $2^{16}$ possible sequences. After 
8000 MC sweeps, corresponding to
a few CPU seconds on a DEC Alpha 200, 
all sequences except the correct one have been removed.

\begin{figure}[tb]
\begin{center}
\vspace{-43mm}
\mbox{
  \hspace{-31mm}
  \psfig{figure=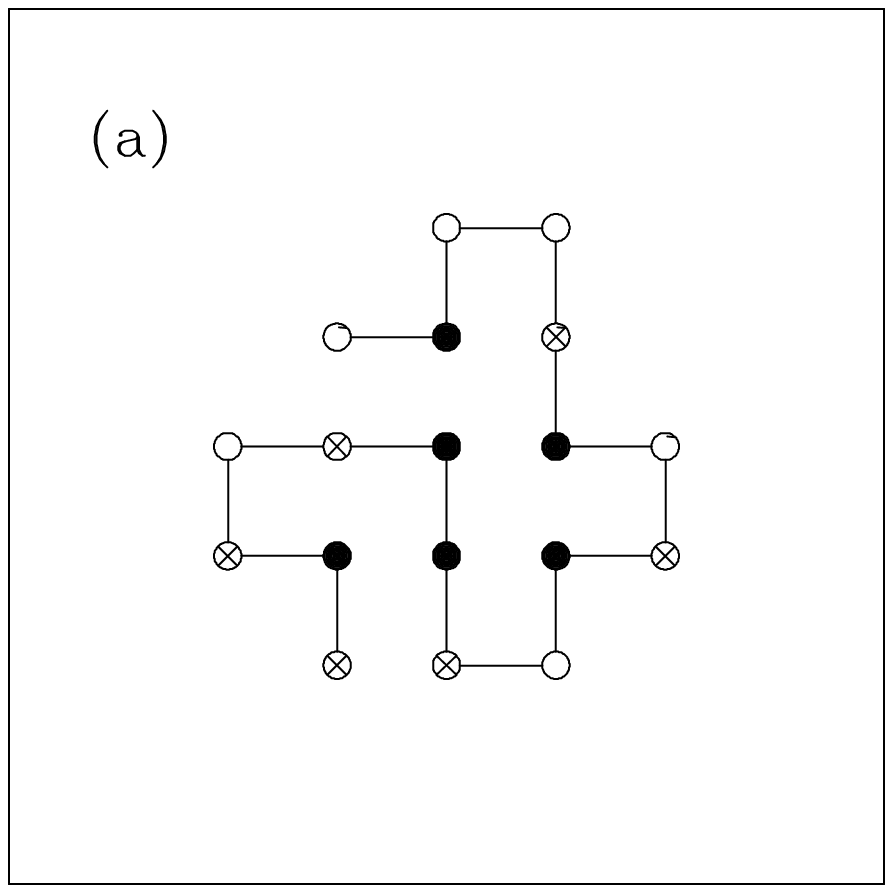,width=10.5cm,height=14cm}
  \hspace{-30mm}
  \psfig{figure=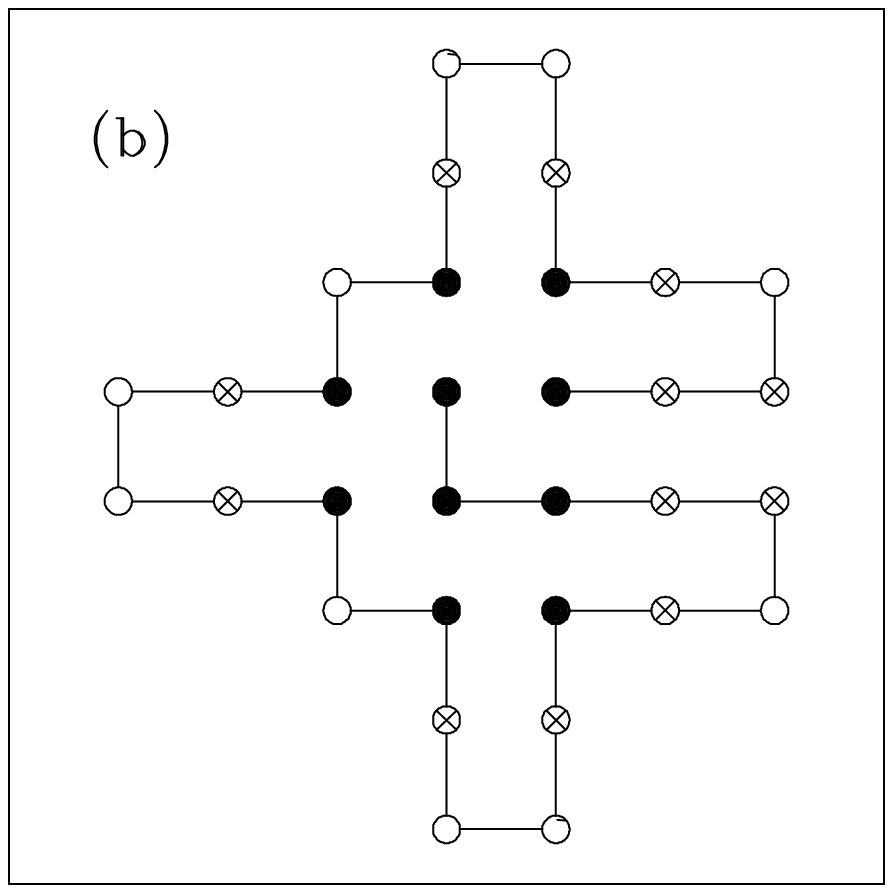,width=10.5cm,height=14cm}
}
\vspace{-48mm}
\end{center}
\caption{Target structures for (a) $N=18$ and (b) $N=32$. Symbols are explained
in the text. The best sequence found for (b) is: 
{\tt HHPPHHPPPPHPHPPPPHPHPPPPHHPPHHHH}.}
\label{fig:2}
\end{figure}

\begin{figure}[tbp]
\begin{center}
\vspace{-43mm}
\mbox{
  \psfig{figure=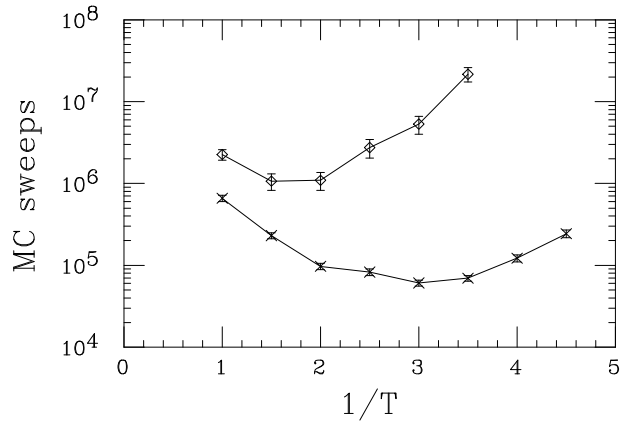,width=10.5cm,height=14cm}
}
\vspace{-43mm}
\end{center}
\caption{The number of MC sweeps needed to single out the
seven sequences in Table~\protect\ref{tab:1}.
Each data point is an average 
over 50 experiments, each started from the full set of all $2^{18}$ 
sequences. Shown are both 
results obtained with ($\times$) and without ($\diamond$) stochastic 
sequence moves.}    
\label{fig:3}
\end{figure}

Next, we turn to the $N=18$ target structure shown in Fig.~\ref{fig:2}a.
The seven sequences listed in Table~\ref{tab:1} were found by the
design procedure.  Non-degeneracy is tested and confirmed as in the 
$N=16$ case above.
\begin{table}[h]
\begin{center}
\begin{tabular}{ccc} 
Sequence & MC & exact  \\
\hline 
PHPPPHPPHPPHHPPPHP & 0.2101(08)  & 0.2112 \\
PHPPHHPPHPPHHPPPHP & 0.0625(09)  & 0.0617 \\
PHPPPHPPHPHHHPPPHP & 0.3104(18)  & 0.3113 \\
PHPPPHPPHPPHHHPPHP & 0.0495(03)  & 0.0495 \\
PHPPPHPHHPPHHHPPHP & 0.1765(19)  & 0.1757 \\
PHPPPHPPHPPHHHPHHP & 0.1102(07)  & 0.1110 \\
PHPPPHPPHPHHHHPPHH & 0.0807(19)  & 0.0797 \\
\end{tabular}
\end{center}
\caption{$P(\sigma)$ for those seven $N=18$ sequences that 
design the structure shown in Fig.~\protect\ref{fig:2}a ($T=1/3$). 
Listed are both the results from our multisequence simulation (MC) and 
the exact results, obtained by enumeration.}
\label{tab:1}
\end{table}                                                             
The multisequence simulation is then continued with only the seven
sequences left in order to estimate $P(\sigma)$ and $P(r_0|\sigma)$.
As can be seen from Table~\ref{tab:1}, the results agree very well
with the exact results. Since all seven sequences have $r_0$ as their
unique ground state, $P(\sigma)$ is constant for $T=0$. 
At $T=1/3$, the temperature studied here, the $P(\sigma)$'s are not
perfectly equal, but similar enough to allow for good mobility in
sequence space. In summary, our method has removed those sequences
that do not have $r_0$ as their unique ground state, and it also
provides $P(r_0|\sigma)$ for all the surviving sequences.
Note that the remaining sequences all have the
same monomer type at 12 of the 18 positions. These are marked by
filled (H) and open (P) circles in Fig.~\ref{fig:2}a.
In the second part of our simulations, where $P(r_0|\sigma)$ is
estimated, it is clear that the stochastic sequence moves are
essential. How useful these moves are in the first part, the elimination
process, is less clear. To investigate this, we performed calculations 
both with and without these moves.
In the latter case the simulated sequence 
is replaced only if it is to be removed from the simulation, and 
is then replaced by a randomly chosen sequence among the remaining ones.  
In Fig.~\ref{fig:3} we show the number of MC sweeps needed to 
remove all sequences except those in Table~\ref{tab:1}, as
a function of $1/T$.
The results show that the required number of
sweeps can be reduced by more than a factor of 10 by adding the stochastic
sequence moves. Furthermore, the efficiency is less $T$ dependent.
The cost of the sequence moves is negligible.

We next turn to the $N=32$ target structure shown in Fig.~\ref{fig:2}b,
which is designed by hand since exhaustive enumeration is impracticable
for this problem size. 
It is readily verified that this structure represents 
the minimum energy for the sequence with H at the filled circles and P at 
all the other positions along the chain (see Fig.~\ref{fig:2}b). 
As with any other method, it is for large chains not 
feasible to explore the entire sequence space with our multisequence method.
However, as in the $N=18$ example above,
a given structure typically exhibits several positions where
$\sigma_i$ is effectively frozen to H or P 
(see e.g.~\cite{Yue:95,Li:96}).
It turns out that such positions can be easily detected by means of a    
trial run. This leads us to a two-step bootstrap procedure.

\begin{figure}[tbp]
\begin{center}
\vspace{-43mm}
\mbox{
  \psfig{figure=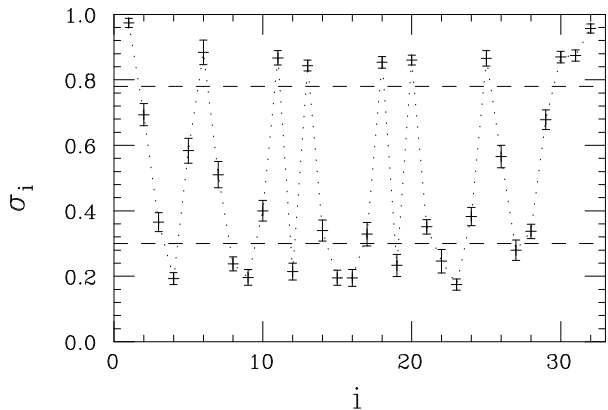,width=10.5cm,height=14cm}
}
\vspace{-43mm}
\end{center}
\caption{Average of $\sigma_i$ against $i$ for the surviving 
sequences from 10 runs, each started with a set of $10^5$ random 
sequences ($N=32$). The upper and lower lines represent $\sigma^{(1)}$ 
and $\sigma^{(2)}$, respectively (see text). The length of each run 
is 20000 MC sweeps (45 CPU seconds) and the number
of surviving sequences varies between 15 and 57.}
\label{fig:4}
\end{figure}

The first step amounts to picking sets of random sequences 
and gauging $\sigma_i$ for the surviving sequences. 
The $\sigma_i$ profile obtained this way for the target 
structure in Fig.~\ref{fig:2}b is shown in Fig.~\ref{fig:4}, 
from which it is clear that indeed many $\sigma_i$ exhibit a 
clear preference for either P or H.
Based on this, we divide the positions along the chain 
into three groups corresponding to $\sigma_i>\sigma^{(1)}$ 
(filled circles in Fig.~\ref{fig:2}b), $\sigma_i<\sigma^{(2)}$ 
(open circles) and $\sigma^{(2)}<\sigma_i<\sigma^{(1)}$ (crosses), as
indicated in Fig.~\ref{fig:4}. We then rerun the algorithm with those 
$\sigma_i$ in the first two groups clamped to H ($\sigma_i=1$) and 
P ($\sigma_i=0$), respectively, and those in the third group left open, 
which corresponds to a set of  $2^{12}$ sequences. In 20 CPU minutes
($5\cdot10^6$ MC sweeps) this set was reduced to 200 sequences. These can be 
readily studied in a final multisequence run, and the best sequence found is 
given in Fig.~\ref{fig:2}. Stability was confirmed by repeating the procedure 
for different random seeds. We also performed runs where the elimination 
process was continued for much longer, which finally contained 167 
surviving sequences. While it could be that this is still not the proper 
asymptotic value, we feel confident that the best sequence is correctly  
identified by our method.

It is clear that this method can be generalized to a corresponding 
multi-step procedure for very large chains. 


In summary, we have developed a new efficient MC method for protein 
design by maximizing conditional probabilities  
using the multisequence method. 
The method circumvents calculations of partition functions by 
a judicious choice of the multisequence sample weights. Large 
chains are feasible with the approach by means of a bootstrap 
procedure that limits the search in sequence space. The method, which 
is successfully explored on two-dimensional lattice models, can 
easily be used in off-lattice models \cite{Irback:96c}.
An alternative to prune $\sigma$-space by
removing $E(r,\sigma)\le E(r_0,\sigma)$ sequences is to 
discard high $P(\sigma)$ sequences (see Eq.~(\ref{DF2})). 
      
\newpage



\end{document}